\documentstyle[prl,aps,twocolumn,epsfig]{revtex}
\draft
\title{Dynamic instability of a rotating Bose-Einstein condensate}
\author{Subhasis Sinha and Yvan Castin}
\address{Laboratoire Kastler Brossel, Ecole normale sup\'erieure,
24 rue Lhomond, 75 231 Paris Cedex 5, France}
\begin{document}
\maketitle
\begin{abstract}
We consider a Bose-Einstein condensate subject to a rotating
harmonic potential, in connection with recent experiments leading to
the formation of vortices. We use the classical hydrodynamic approximation
to the non-linear Schr\"odinger equation to determine almost analytically
the evolution of the condensate. 
We predict that this evolution can exhibit dynamical instabilities, for the 
stirring procedure previously demonstrated at ENS
and for a new stirring procedure that we put forward.
These instabilities take place within the range
of stirring frequency and amplitude for which vortices
are produced experimentally. They provide therefore an initiating
mechanism for vortex nucleation.
\end{abstract}
\pacs{03.75.Fi, 05.30.Jp }

Quantized vortices in superfluid helium II,
in particular the issue of vortex nucleation in a rotating container, 
have long been the subject of intense work \cite{Donnelly}. 
With the recent production of gaseous Bose-Einstein condensates
\cite{revue} the subject has gained a renewed interest.
On the experimental side, three groups have succeeded in obtaining vortices
in atomic condensates, with two different techniques:
a phase imprinting technique at JILA \cite{JILA_vortex} and
the equivalent of the helium rotating bucket experiment at 
ENS \cite{ENS} and at MIT \cite{MIT}.
At ENS a rotating laser beam superposed onto the magnetic trap
holding the atoms creates a harmonic rotating potential
with adjustable anisotropy $\epsilon$ and rotation frequency $\Omega$.
For a well chosen range of variation for $\Omega$ 
one or several vortices are nucleated, and then detected as holes
in the density profile of the gas after ballistic expansion \cite{ENS}
or by a measurement of the angular momentum of the condensate \cite{ENS_lz}.
A striking feature of the ENS experimental results 
is that, for a very weak anisotropy $\epsilon$,
nucleation of vortices takes place 
in a narrow interval of rotation frequencies $[\Omega_{\mbox{\scriptsize min}},
\Omega_{\mbox{\scriptsize max}}]$
around $0.7\omega_\perp$,
where $\omega_\perp$ is the mean oscillation frequency 
of the atoms in the $x-y$ plane,
whatever the number of atoms or the oscillation frequency $\omega_z$
along $z$ in the experiment \cite{Houches}.

While the JILA experiment is well understood theoretically \cite{Holland}
the situation is more involved for the ENS experiment.
Several theoretical articles, inspired by the case of superfluid helium,
have tried to predict the value of the lower vortex nucleation frequency
$\Omega_{\mbox{\scriptsize min}}$ from purely thermodynamic arguments
\cite{Baym_and_Pethick,Stringari_and_Dalfovo,Sinha,Pethick_again,Fetter,Castin_Dum,Machida}.
The proposed values for $\Omega_{\mbox{\scriptsize min}}$
are significantly different from the observed value of $0.7
\omega_\perp$, or depend 
on the trap aspect ratio $\omega_z/\omega_\perp$ or on the atom
number, in contradiction with the observations at ENS. 
Also thermodynamical reasonings are not able to predict the 
upper vortex nucleation frequency $\Omega_{\mbox{\scriptsize max}}$, 
which is also close to $0.7\omega_\perp$ for low anisotropy $\epsilon$.

In this paper we consider the time dependent problem of a condensate
subject to a harmonic stirring potential.
We use the classical 
hydrodynamic approximation to the time dependent 
Gross-Pitaevskii equation (GPE),
an approximation well justified for the ENS parameters 
\cite{Stringari_modes}. 
We are then able to reformulate
the partial differential hydrodynamic equations 
in terms of ordinary differential
equations, which allows an almost analytical solution \cite{obstacle}.
Our main result is the discovery of dynamical instabilities in the evolution
of the condensate for a certain range of the rotation frequency and of the trap
anisotropy. These instabilities will invalidate the classical hydrodynamic
approximation after some evolution time. We have checked
with a numerical solution of the Gross-Pitaevskii equation that
vortices then enter the condensate.

The existence of such a dynamical instability 
explains why in earlier numerical work the time dependent
Gross-Pitaevskii equation was found to nucleate vortices
\cite{Fetter,Burnett,Feder}.
Furthermore 
the instability range that we predict is very close to the 
experimentally observed range
of vortex nucleation, for various stirring procedures.
For the stirring procedure of \cite{ENS,ENS_lz}
we recover the \lq\lq universal"
numerical value $0.7$ for $\Omega/\omega_\perp$ leading to vortex
nucleation for low anisotropies $\epsilon$.
We provide a simple physical interpretation of this value:
for $\Omega/\omega_{\perp}=1/\sqrt{2}\simeq 0.7$ the 
harmonic stirring potential 
resonantly excites a quadrupole mode of the condensate \cite{note}, which
induces
large oscillations of the condensate and eventually a dynamical instability
sets in.
We also investigate a new excitation procedure to nucleate vortices, 
that has recently been implemented at ENS \cite{ENS_companion}: the rotation
of the stirring potential is set up very slowly. The gas then follows adiabatically
a branch of steady state until the branch becomes dynamically unstable. The corresponding
lower rotation frequency $\Omega$ that we predict for
vortex nucleation is also very close to the experimental
value.

In our model atoms are trapped in a harmonic potential rotating at
the instantaneous frequency $\Omega(t)$ around $z$ axis.
For convenience all the calculations of this
paper are done in a rotating frame where the trap axes are fixed.
The trapping potential then reads:
\begin{equation}
U(\vec{r}, t) = \frac{1}{2}m \omega_{\perp}^{2}\left\{[1 - \epsilon(t)]x^{2} +
[1 + \epsilon(t)]y^{2} +\left(\frac{\omega_{z}}{\omega_{\perp}}\right)^{2}z^{2}\right\}
\end{equation}
where $m$ is the mass of an atom, $\epsilon(t)$ is the trap anisotropy at time $t$.
The parameters $\omega_{\perp}$ and $\omega_{z}$ 
are the oscillation frequencies of the atoms in transverse and axial directions for vanishing
anisotropy of the stirring potential.
Within the mean field approximation,
the time evolution of the condensate field or macroscopic wavefunction
$\psi(\vec{r},t)$ can be
described by the time dependent Gross-Pitaevskii equation
\cite{Review_Stringari}:
\begin{equation}
\label{eq:gpe}
i\hbar \frac{\partial \psi}{\partial t} = \left[-\frac{\hbar^{2}}{2
m} \vec{\nabla}^{2} + U(\vec{r}, t) + g |\psi|^{2} - \Omega(t)
\hat{L}_{z}\right] \psi,
\end{equation}
where $g = 4 \pi \hbar^{2} a/m$ is the coupling
constant, proportional to the $s$-wave scattering
length $a$ of the atoms, here taken to be positive, and where the inertial term 
proportional to the angular 
momentum operator $\hat{L}_z$ along $z$-axis accounts for the frame rotation.
The condensate field $\psi$ can be written in terms of
density $\rho$ and phase $S$,
\begin{equation}
\psi(\vec{r}, t) = \sqrt{\rho(\vec{r}, t)} e^{i S(\vec{r}, t)/\hbar}.
\end{equation}
The equation obtained from the GPE for $\rho$ is just the continuity equation.
The equation for $S$ contains the so-called quantum pressure term $\hbar^2\vec{\nabla}^2
\sqrt{\rho}/2m\sqrt{\rho}$ that we neglect here as compared to the mean-field
term $\rho g$ in the Thomas-Fermi approximation. We obtain:
\begin{eqnarray}
\label{eq:evol_rho}
\frac{\partial \rho}{\partial t} & = & 
-\mbox{div}\,\left[
\rho\left(\frac{\vec{\nabla} S}{m}-\vec{\Omega}(t)\times \vec{r}\right)\right] \\
-\frac{\partial S}{\partial t} & = & \frac{(\vec{\nabla} S)^2}{2 m} +
U(\vec{r}, t) + g \rho - (\vec{\Omega}\,(t) \times \vec{r}\,)\cdot
\vec{\nabla} S.
\label{eq:evol_s}
\end{eqnarray}
A very fortunate feature of the harmonic trap is that
these superfluid hydrodynamic equations can be solved exactly
for a condensate initially at equilibrium in the non-rotating trap
with the following quadratic ansatz for the condensate density and phase
\cite{cut_for_rho}:
\begin{eqnarray}
\label{eq:ansatz_rho}
\rho_c(\vec{r}, t) & = & \rho_0(t) +
 \frac{m \omega_{\perp}^{2}}{g}\sum_{i,j = 1}^{3} x_{i} A_{ij}(t) x_{j},\\
S_c(\vec{r}, t) & = & s_{0}(t) + m \omega_{\perp}
\sum_{i,j = 1}^{3}x_{i} B_{ij}(t) x_{j},
\label{eq:ansatz_s}
\end{eqnarray}
where $x_{1}$, $x_{2}$ and $x_{3}$ are the coordinates along $x$, $y$ and $z$
axes respectively. The time dependent dimensionless coefficients $A_{ij}$ and $B_{ij}$ form
$3 \times 3$ symmetric matrices $A$ and $B$ which from
Eqs.(\ref{eq:evol_rho},\ref{eq:evol_s}) obey the evolution equations:
\begin{eqnarray}
\label{eq:Adot}
\omega_\perp^{-1}\frac{d A}{dt} & = & -2 A\ \mbox{Tr} B - 2 \{A, B\} + \frac{\Omega}{\omega_\perp} 
[R, A], \\
\label{eq:Bdot}
\omega_\perp^{-1}\frac{d B}{dt} & = & -2 B^{2} - W - A + \frac{\Omega}{\omega_\perp} [R, B]
\end{eqnarray}
where $\{,\}$ stands for the anti-commutator, $[,]$ stands for
the commutator of two matrices,
the matrix $W$ is diagonal, with components
$W_{11}= (1 - \epsilon (t))/2$, $W_{22} = (1+\epsilon(t))/2$, and $W_{33} =
(\omega_{z}/\omega_{\perp})^{2}/2$, and the matrix $R$, originating from the vectorial
product
in $\hat{L}_z$, has vanishing elements except for $R_{12}=
-R_{21}=1$ \cite{Olshanii}. Note that these equations do not depend on the number
of atoms nor on the coupling constant $g$.

\begin{figure}[htb]
\hskip -10mm
\vspace*{-3mm}
\begin{center}
\epsfig{file=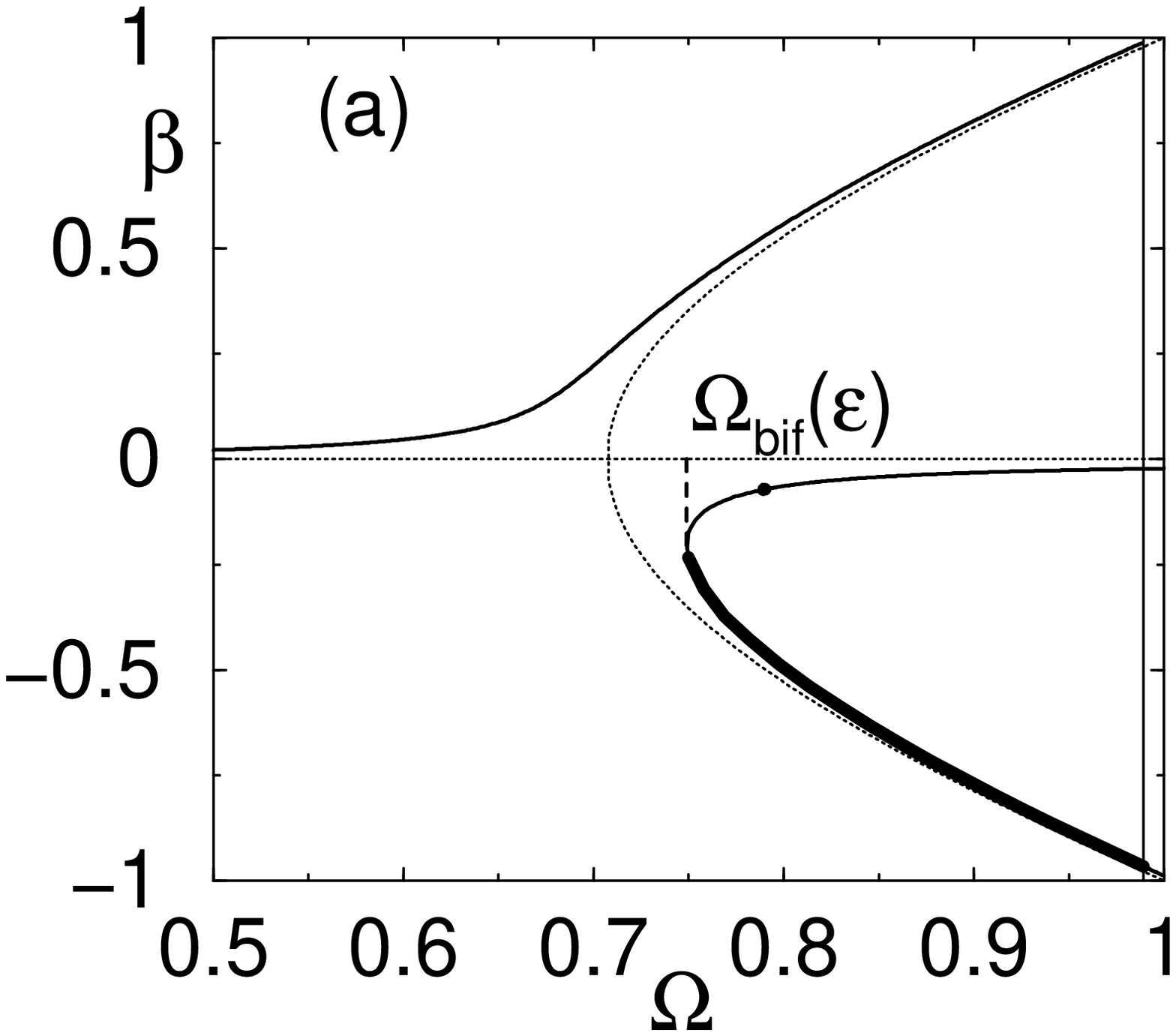,height=3.84cm}
\epsfig{file=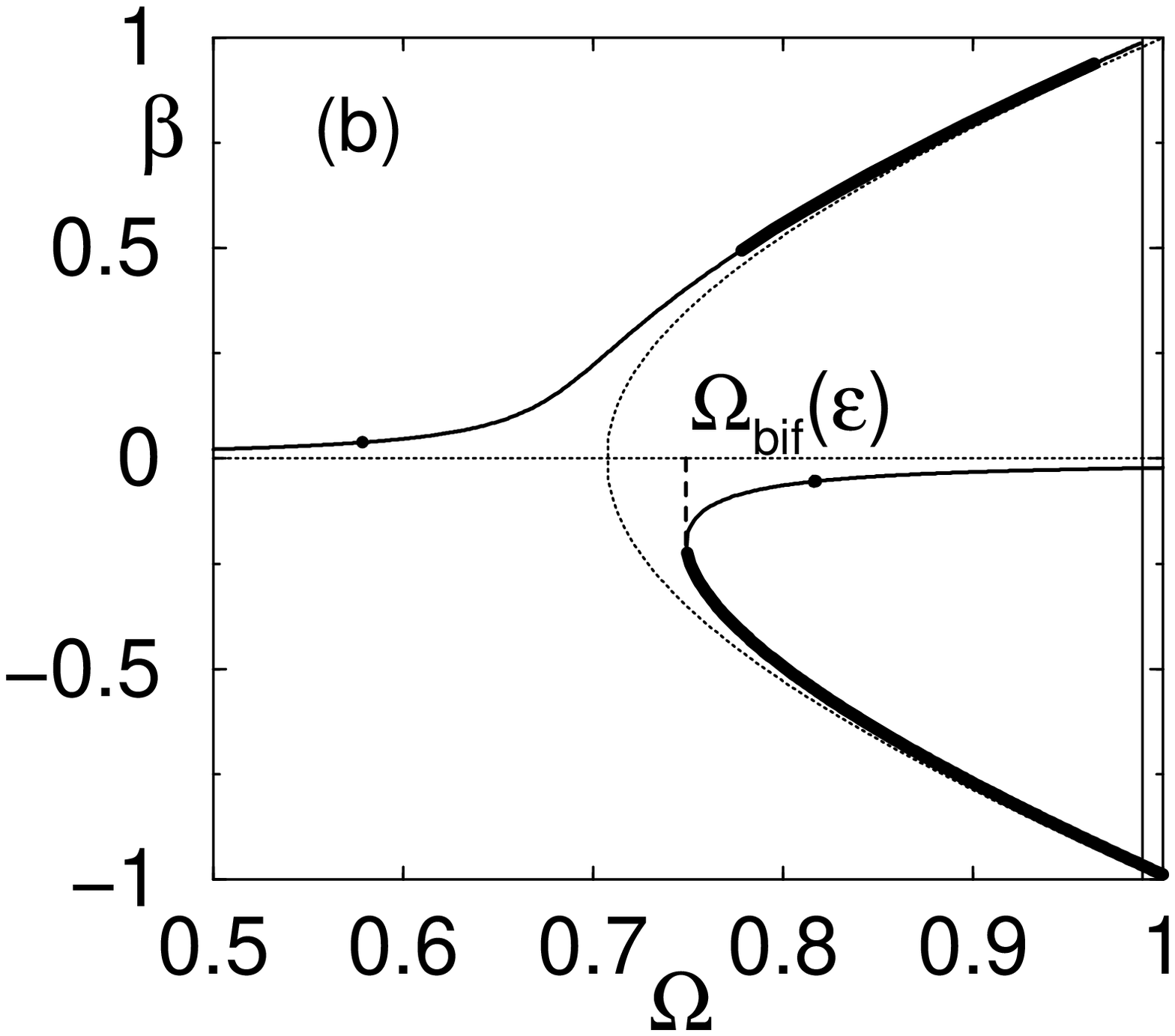,height=3.84cm}
\end{center}
\caption{
\label{fig:beta}
Phase parameter $\beta$ for a steady state condensate
as function of the rotation frequency $\Omega$ in units of $\omega_\perp$.
Dotted lines: $\epsilon=0$. Solid lines: $\epsilon =0.022$. For
$\epsilon=0.022$ and
$\omega_{z}/\omega_{\perp} =0.1$ 
the thick lines on the curves indicate
where the solution has a dynamical instability (a) of degree $n=2$ and (b) of degree $n=3$.
The vertical line is the border of the center of mass
instability domain for $\epsilon=0.022$.}
\end{figure}

In a first stage, 
it is very important to study steady
state solutions of the above equations. We restrict to solutions that have the
same symmetry as the initial state: 
even parity along $z$, this parity being preserved by time evolution.
We then find an unique class of solutions, reproducing the results of 
\cite{Stringari}: the condensate phase
varies as $S(\vec{r}\,)  =  m \omega_\perp \beta x y$
where $\beta$ is a real root of
\begin{equation}
\beta^{3} + 
\left(1 -2 \frac{\Omega^{2}}{\omega_{\perp}^{2}}\right)
\beta -\frac{\Omega}{\omega_\perp}
\epsilon = 0.
\end{equation}
The steady state matrix $A$ is diagonal with elements given in 
\cite{Stringari}.
We have plotted in figure~\ref{fig:beta} the values of $\beta$ as function of the rotation
frequency $\Omega$ for a fixed anisotropy $\epsilon$. For $\Omega$ between zero
and a bifurcation value $\Omega_{\mbox{\scriptsize bif}}(\epsilon)$ depending on  $\epsilon$
there is a single branch of solution for $\beta$. This branch is supplemented
by two extra branches when $\Omega > \Omega_{\mbox{\scriptsize bif}}(\epsilon)$.

We now turn back to the time dependent problem.
Clearly a condensate with a vortex cannot
be described within the quadratic ansatz (\ref{eq:ansatz_rho},\ref{eq:ansatz_s})
as the phase $S_c$ corresponds to an irrotational velocity flow.
The actual scenario for the vortex nucleation that we put forward is the following: initially very
small deviations $\delta\rho(\vec{r},t)$ of the condensate density 
and $\delta S(\vec{r},t)$ of the condensate phase from the quadratic shapes
$\rho_c$ and $S_c$ may grow exponentially fast in the course of time
evolution, eventually leading the condensate to a structure very different from 
Eqs.(\ref{eq:ansatz_rho},\ref{eq:ansatz_s}). This may happen when a dynamical instability
is present.

To reveal such an instability we obtain from the evolution equations (\ref{eq:evol_rho},
\ref{eq:evol_s}) linearized equations of motion for initially
small deviations $\delta\rho$ and $\delta S$ from $\rho_c$ and $S_c$:
\begin{eqnarray}
\frac{D \delta \rho}{D t} & = & -\mbox{div}\,
\left(\rho_c \frac{\vec{\nabla}\delta S}{m}\right)-\delta \rho \frac{\vec{\nabla}^{2} S_c}{m} , 
\label{delta_rho}\\
\frac{D \delta S}{D t} & = & -g \delta \rho. \label{delta_s}
\end{eqnarray}
In these equations, we have introduced the convective derivative  $\frac{D}{Dt} \equiv 
\frac{\partial}{\partial t} + \vec{v}_{c}(\vec{r},
t)\cdot \vec{\nabla}$ where $\vec{v}_{c}=\vec{\nabla}S_c/m - \vec{\Omega} \times
\vec{r}$
is the velocity field of the condensate in the rotating frame.
A polynomial ansatz for $\delta S$ and $\delta \rho$ of an arbitrary total degree $n$
in the coordinates $x$, $y$ and $z$ solves these linear equations exactly
\cite{more_on_degree}.
This is another nice consequence of the harmonicity of the trap.
In practice, we calculate the evolution operator ${\cal U}_n(t)$
mapping the coefficients of the polynomials at time zero onto their values 
after a time evolution $t$.
Dynamical instability takes place when one or several eigenvalues 
of ${\cal U}_n$  grow exponentially fast with time $t$.
Note that after rescaling of the variables, Eqs.(\ref{delta_rho},\ref{delta_s}) become
independent of the number of atoms and of the coupling constant $g$, in a way
similar to Eqs.(\ref{eq:Adot},\ref{eq:Bdot}).

Now we perform a linear stability analysis
for two different stirring procedures of
the condensate.

{\bf Procedure I:} The ellipticity
of the stirring potential $\epsilon$ is kept fixed and the
rotation frequency $\Omega(t)$ of the stirrer is very slowly
ramped up from zero to its final value. The condensate, initially
in steady state with a vanishing parameter $\beta$, 
adiabatically follows the
upper branch of steady states with $\beta\ge 0$,
see figure~\ref{fig:beta}.
It is then sufficient to determine the dynamic stability of the upper branch
of steady states. This greatly simplifies the calculation as one just has
to identify eigenmodes of Eqs.(\ref{delta_rho},\ref{delta_s}) evolving in
time as $\exp(\lambda t)$ where the eigenvalue $\lambda$ is a complex number.
Dynamical instability takes place when $\lambda$ can have a strictly positive
real part.
As shown in figure~\ref{fig:beta} for $\epsilon=0.022$ we find that the upper branch for $\beta$
is stable for modes of degree $n=2$ but presents two instability intervals for the modes
of degree $n=3$, a very narrow interval around $\Omega=0.58\omega_\perp$ 
and a broader interval starting at $\Omega=0.778\omega_\perp$.

\begin{figure}[htb]
\hskip -10mm
\vspace*{-3mm}
\begin{center}
\epsfig{file=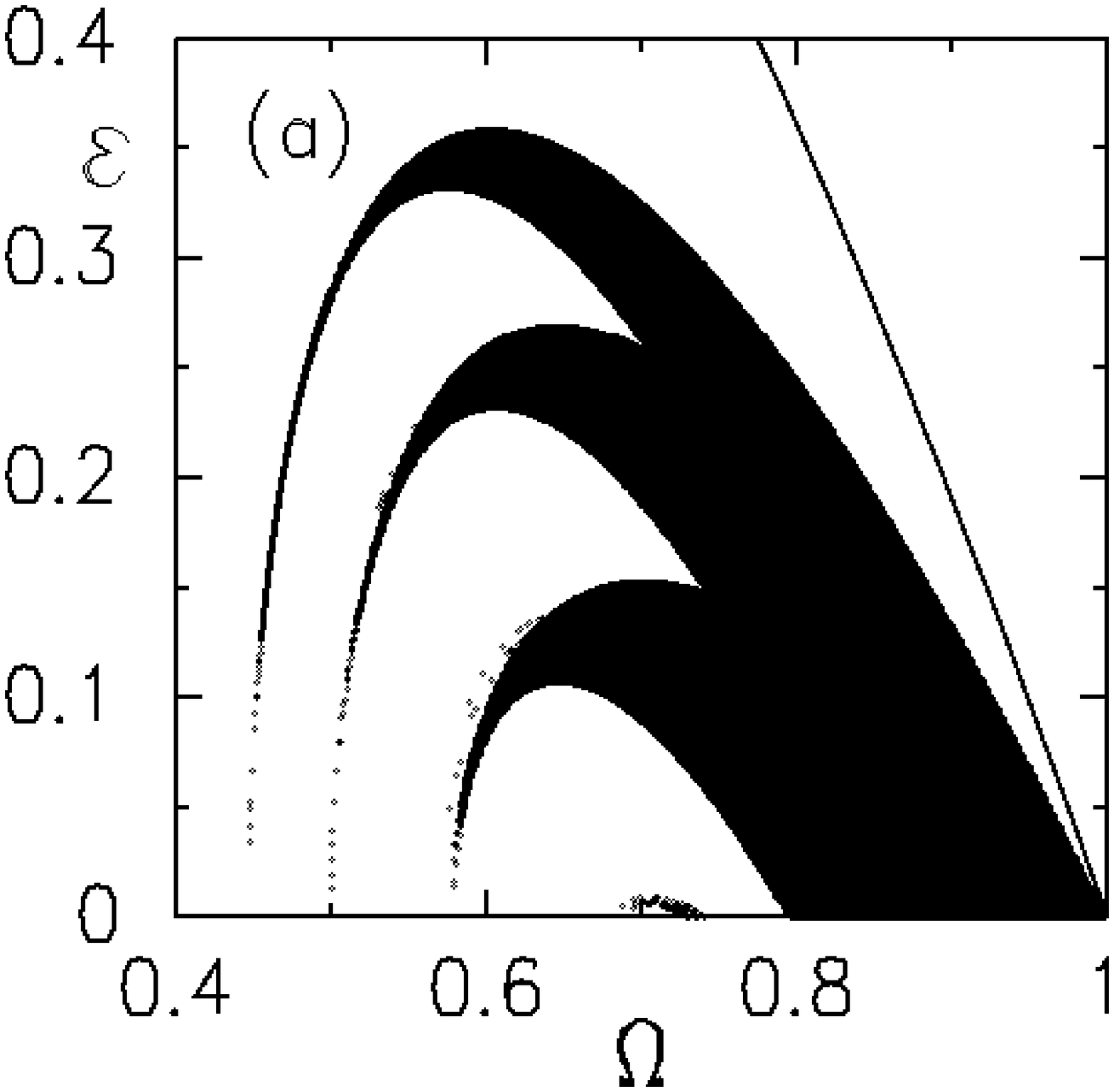,height=3.8cm}
\epsfig{file=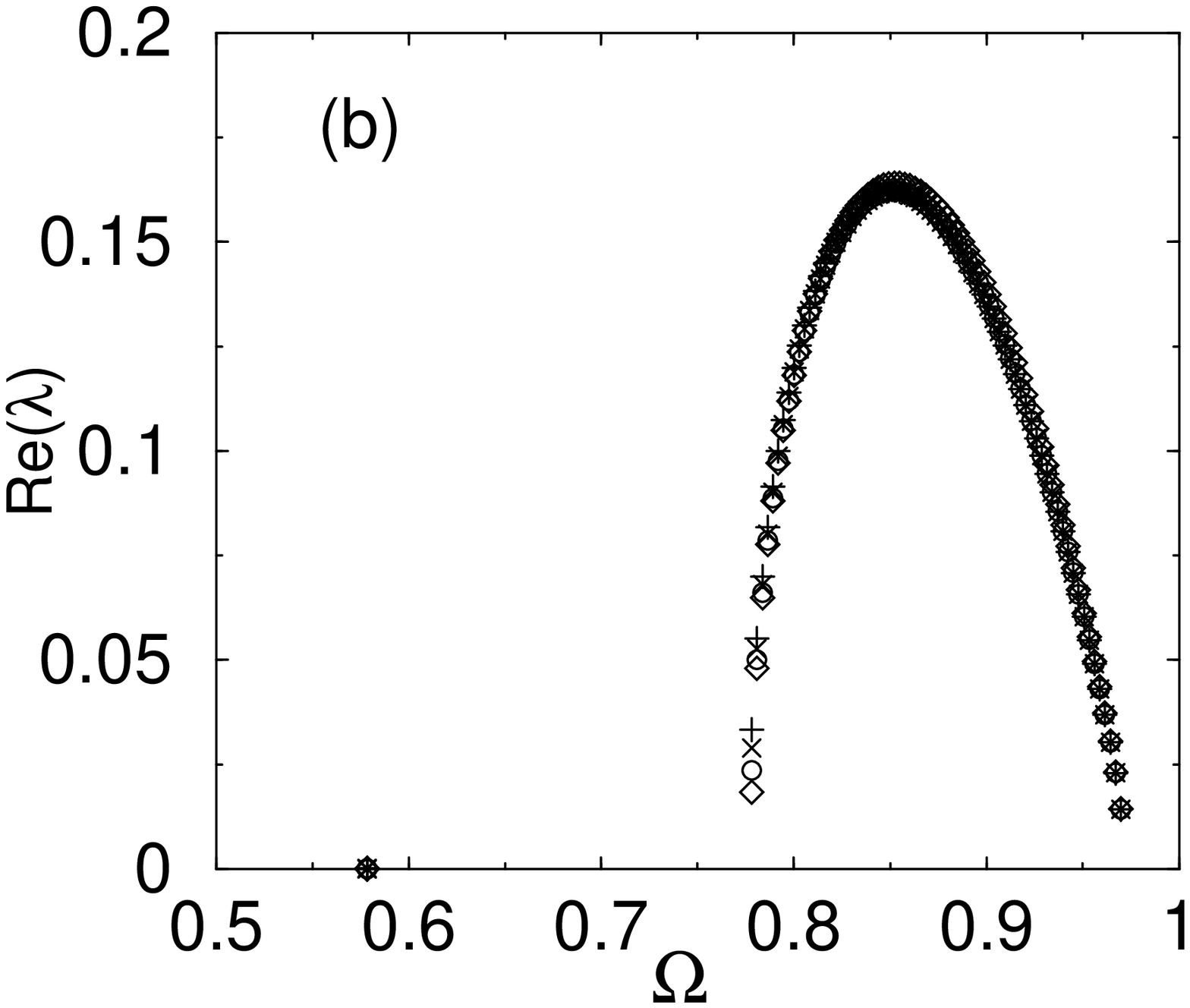,height=3.8cm}
\end{center}
\caption{
\label{fig2}
For the upper branch of steady state condensates: (a) 
Dark areas: instability domain in the $\Omega-\epsilon$ plane for $\omega_z/\omega_\perp
=0.1$ 
for degrees $n$ equal to 3, 4 and 5 (crescents from bottom to top). 
There is no dynamical instability for $n\leq 2$. Solid line: border
$\Omega^2= (1-\epsilon)\omega_\perp^2$ of the branch existence domain. 
(b) Maximal instability exponent Re$(\lambda)$
for $n=3$ as function of $\Omega$, for $\epsilon=0.022$, and $\omega_z/\omega_\perp=
0.1$ ($\diamond$), 0.5 ($\circ$), 1.0 ($\times$) and $\sqrt{8}$ ($+$). $\Omega$, Re($\lambda$)
are in units of $\omega_\perp$.}
\end{figure}

We have investigated in a systematic way the instability range of the
upper branch of steady states, by varying the anisotropy $\epsilon$, the rotation
frequency $\Omega$ and the degree $n$.
The instability domain in the $\Omega-\epsilon$ plane for modes of degree
3 is mainly made
of a crescent,
and the inclusion of higher degree modes ($n = 4, 5$)
add extra crescents from above, see figure~\ref{fig2}a. 
Each crescent has on the $\epsilon=0$ axis (i) a broad basis at 
$\Omega>\omega_\perp/\sqrt{2}$, with a non-zero instability exponent,
and (ii) a very narrow edge at 
$\Omega=\omega_{\perp}/\sqrt{n}<\omega_{\perp}/\sqrt{2}$ 
with a vanishing instability exponent.
We show in figure~\ref{fig2}b that the maximal instability exponent for $n=3$
has a remarkably weak dependence on $\omega_{z}/\omega_{\perp}$.

{\bf Procedure II:} This is the original experimental
scenario of \cite{ENS,ENS_lz}, where the stirring potential is rotated at a fixed
frequency and the ellipticity of the stirrer is turned on
from zero to its final value $\epsilon_f$ abruptly.
In this case we cannot rely on adiabatic following for the condensate
density and phase, so that we solve the time dependent equations 
(\ref{eq:Adot},\ref{eq:Bdot}) for
$\rho_{c}(t)$ and $S_{c}(t)$.
Then we perform a linear stability analysis as discussed above: 
we evolve a generic polynomial
ansatz of degree $n$
for the fluctuations $\delta\rho$ and $\delta S$ according to
Eqs.(\ref{delta_rho},\ref{delta_s}), which allows to 
construct the evolution operator ${\cal U}_n(t)$ and to calculate
$Z_{\mbox{\scriptsize max}}(t)$, the eigenvalue of ${\cal U}_n(t)$
with the largest modulus. Then we define the mean instability exponent
Re\,$\langle\lambda\rangle$ as the mean slope of 
$\ln|Z_{\mbox{\scriptsize max}}(t)|$ as function of time.

\begin{figure}[htb]
\begin{center}
\hskip -7mm
\epsfig{file=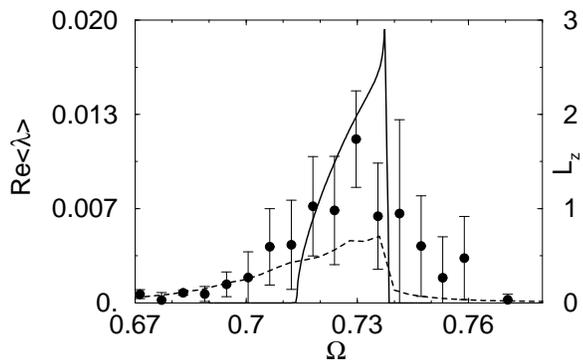,height=5cm}
\end{center}
\caption{
\label{fig3}
For the stirring procedure II, with $\epsilon_f=0.01$ and 
$\omega_z/\omega_\perp=0.1$:
Solid line:
mean instability exponent Re\,$\langle\lambda\rangle$ (see text)
of the vortex free classical hydrodynamic solution $\rho_c, S_c$
as function of $\Omega$,
for $n=3$. 
Dashed line: mean angular momentum per particle $\bar{L}_z$
(see text) obtained from $\rho_c,S_c$. 
Filled disks: experimentally measured
angular momentum $L_z$ per particle in the condensate after
vortices have possibly entered the condensate \protect\cite{Houches}. 
The initial steady state condensate
in the calculation of $\bar{L}_z$ has a chemical
potential $\mu=10\hbar\omega_\perp$,
close to the experimental value.
Re\,$\langle\lambda\rangle$ and $\Omega$ are in units of $\omega_\perp$,
and $L_z$, $\bar{L}_z$ are in units of $\hbar$.}
\end{figure}

This reveals that within certain range of rotation
frequency the system becomes dynamically unstable, see the
solid line in figure~\ref{fig3}. 
In the limit of a low anisotropy $\epsilon$
the instability sets in when the rotation frequency $\Omega$ is close to the
value $\simeq 0.7 \omega_{\perp}$: 
in the lab frame,
the stirring potential of frequency $2\Omega$
is then resonant with a quadrupole mode 
\cite{Stringari_modes,Stringari_and_Dalfovo2}
of the condensate of frequency $\sqrt{2}\omega_\perp$,
and induces large amplitude oscillations
of the condensate, resulting in a dynamical instability.  
More precisely, the condensate
described by the quadratic ansatz $\rho_c,S_c$
has an angular momentum oscillating in time
around a non-zero value $\bar{L}_z$. 
This value $\bar{L}_z$
is a peaked function of the rotation frequency $\Omega$, 
shown in dashed line in figure 3. 
The peak of $\bar{L}_z$ is not exactly located
at $\Omega=0.7\omega_\perp$ 
because of non-linear effects in Eqs.(\ref{eq:Adot},
\ref{eq:Bdot}). 
The peak structure of the instability
exponent in figure 3 is alike the peak structure
of $\bar{L}_z$, with a narrower width as dynamical
instability of the vortex free solution $\rho_c, S_c$ sets in for the
higher values of $\bar{L}_z$ only.
For values of $\Omega$ significantly above
or below 0.7$\omega_\perp$ 
the stirrer is out of resonance with the
quadrupole mode and induces only
small and stable oscillations of the condensate.
For larger values of $\epsilon$, the instability interval in
$\Omega$ broadens.
We have also checked that 
the instability interval depends weakly on $\omega_z/\omega_\perp$.

What is the connection between the dynamical instabilities found here and 
the nucleation of vortices ?
To obtain a theoretical answer to this question, one
has to go beyond a linear stability
analysis to determine the evolution of the condensate in the
long run: for a few values of the rotation frequency $\Omega$
and for procedures I and II,
we have checked by a numerical integration of the time dependent
GPE in three dimensions, that vortices are indeed nucleated
in the predicted instability domains:
after some evolution time,
the angular momentum in the numerical solution suddenly becomes larger than
the classical hydrodynamic prediction, as vortices enter in the condensate.
An experimental answer to this question for the stirring procedure I 
has been provided recently at
ENS \cite{ENS_companion}: the first clear evidence of a 
vortex appears for a rotation frequency
$\Omega=0.77\omega_{\perp}$, very close to our prediction $\Omega=0.778\omega_\perp$.
The agreement is also very good for procedure II as shown in
figure~\ref{fig3}: our instability domain in $\Omega$ coincides with the experimental
vortex nucleation interval within a few percent.

In summary, the dynamical instabilities that we have identified 
provide  an initiating mechanism for the production of
vortices in a condensate stirred by a harmonic potential, in excellent
agreement with the experimental results at ENS. 

We thank S. Rica, V. Hakim, G. Shlyapnikov, F. Chevy, 
K. Madison and J. Dalibard for helpful discussions.
We acknowledge financial support from Minist\`ere de la Recherche et de
la Technologie.
LKB is a unit\'e de recherche de l'Ecole normale sup\'erieure et de l'Universit\'e
Pierre et Marie Curie, associ\'ee au CNRS.


\begin{thebibliography}{99}
\bibitem{note}
At MIT the gas was stirred 
with a strongly anharmonic potential \cite{MIT} which excites modes
other than quadrupole and broadens the nucleation frequency interval.
\bibitem{cut_for_rho} 
$\rho_c$ is set to zero where the ansatz is negative.
\bibitem{more_on_degree}
We take `degree' of the modes in the strict sense: 
the degree of the highest degree
monomial with non vanishing coefficient.
\end{thebibliography}
\end{document}